# Slice-Connection Clustering Algorithm for Tree Roots Recognition in Noisy 3D GPR Data

Wenhao Luo[(1)], Yee Hui Lee[(1)], Lai Fern Ow[(2)], Mohamed Lokman Mohd Yusof[(2)] and Abdulkadir C. Yucel[(1)]

(1) School of Electrical & Electronic Engineering, Nanyang Technological University, Singapore
({wenhao.luo, eyhlee, acyucel}@ntu.edu.sg)
(2) Centre for Urban Greenery & Ecology, National Parks Board, Singapore
({genevieve_ow, mohamed_lokman_mohd_yusof}@nparks.gov.sg)

*Abstract*—3D mapping of tree roots is a popular ground-penetrating radar (GPR) application. In real field tests, the recognition of tree roots suffers due to noisey reflection patterns from subsurface targets that are not of interest, such as rocks, cavities, soil unevenness, etc. A Slice-Connection Clustering Algorithm (SCC) is applied to separate the regions of interest from each other in a reconstructed 3D image. The proposed method can successfully recognize the radar signatures of the roots and distinguish roots from other objects. Meanwhile, most noise radar features are ignored through our method. The final 3D mapping of the radargram obtained by the method can be used to estimate the location and extension trend of the tree roots. The effectiveness of the proposed system is tested on real GPR data.

## I. Introduction

Ground-penetrating radar (GPR) has been widely used in the detection and reconstruction of underground objects such as pipes and cables [1] because it is a portable and non-destructive technology. In recent years, GPR has been investigated in the tree roots detection area [2].

B-scan images have been researched in depth in some previous studies. In [3], researchers proposed a column-connection clustering (C3) algorithm to cluster regions of interest and to further recognize and fit hyperbola features in the B-scan image when the knowledge of the medium is unknown. In [4], authors made a survey on the identification of hyperbola with prior knowledge of the medium. However, the study on clustering reflection features and fitting objects in processed 3D radar images has not been done.

In this paper, we propose a method named slice-connection clustering (SCC) algorithm to cluster reflection features from the same object together in the 3D preprocessed GPR images. With the SCC algorithm, the regions of interest (ROI)/non-interest are separated into different clusters. We can extract ROI that we want from noisy underground 3D reconstruction image. The algorithm was applied to a set of real tree-root test GPR data, the effectiveness of the algorithm is explained in the experiment and result part of the paper.

## II. Slice-Connection Clustering Algorithm

In this section, we present the preprocessing procedures and the proposed SCC algorithm.

### A. Preliminary Signal Processors

The raw data were processed through the following standard pre-processing methods [1] [5]: (i) Zero-offset removal, to make the mean value of the GPR A-scan equal to zero. (ii) Time-zero correction, to set the time-zero point for the B-scan data constant, reduce the impact of air gap between antennas and soil surface. (iii) Singular value decomposition (SVD), to reduce the clutter noise. (iv) Frequency-wavenumber (F-K) migration, to transform a hyperbola feature of the target's radargram into a focused image showing the object's true location and size.

After the preprocessing steps above, we took each B-scan as a slice and combine them into a 3D C-scan which includes the mappings of both tree roots and targets that are not of interest. The C-scan then went through the proposed SCC algorithm.

### B. Slice-Connection Clustering Algorithm

After an input C-scan matrix was generated with multiple B-scans after pre-processing, the SCC algorithm was applied to separate the selected regions into different clusters. Two concepts are introduced: *Slice Region* and *Connecting Regions* of two slice regions from adjacent slices.

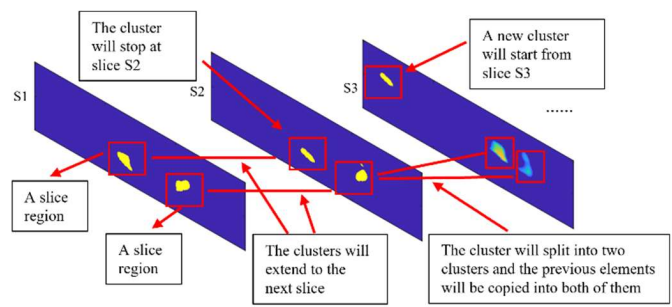

Figure 1. Illustration of the SCC algorithm (detailed in the text).

*Slice Region*: In one slice (a pre-processed B-scan), a focused feature that is transferred from a hyperbolic signature of a target is called a slice region. For example, in Fig. 1, there are two slice regions in Slice S1, Slice S2, three slice regions in Slice S3. The slice regions in each slice are recognized and extracted by applying a clustering method called column-connection clustering (C3) algorithm to 2D slices, the C3 algorithm was a clustering method that was proposed originally to help to fulfill the function of hyperbola recognition and fitting in the denoised GPR B-scan data [3]. The clustering concept of C3 was further adopted here to recognize each focused feature (slice region) of the object. The location of every slice region was then defined by the 2D coordinates in which it appeared.

*Connecting Regions*: If two slice regions from two adjacent slices have pixels from the same 2D coordinates, they are defined as connecting regions. In this paper, we only discuss pixels between two slice regions from adjacent slices. For example, in Fig.1, there are two connecting regions between the first slice S1 and slice S2, and slice S3 have two connecting regions with slice S2.

In Fig.1, if slice S1 is the first slice scanned, then after searching this whole slice, the seeds of two clusters are generated. We call them clusters 1–2 from left to right. Next, Slice S2 is scanned, and the slice regions from this slice are obtained. If two slice regions from adjacent slices have connecting pixels, then the cluster extends to the next slice to include the pixels of the slice region from the later slice. Thus, cluster 1 and cluster 2 are extended to slice S2, and cluster 2 is extended to slice S3. There are two slice regions of slice S3, which are the connecting regions with cluster 2 of slice S2. In this situation, cluster 2 extends to slice S3 and splits into two clusters 2a and 2b. All the pixels in cluster 2 are associated with both clusters with the pixels in the second slice region of slice S3 added to cluster 2a and the pixels in the third slice region of slice S3 added to cluster 2b. As for the first slice region in slice S2, since there is no connecting region in slice S3 with it, cluster 1 stops at slice S2. There is no connecting region in the previous Slice S2 to the first slice region in slice S3, therefore, a new cluster starts from slice S3 with the first slice region as the seeds. This procedure is performed until the last slice is scanned to obtain all the clusters based on slice connection.

### III. EXPERIMENT AND RESULT

In this section, we applied our algorithms to real data sets. The data were collected by a VNA-based bi-statistic common offset SFCW-GPR system [2] [6], the settings of the system followed the ones in [7]. The GPR survey was conducted around a tree which is located along a tree-lined avenue aside from the Chin Bee heavy vehicle parking, Singapore. In the 3D data set, 20 B-scans were included, each B-scan contained 101 A-scan traces with a step of 3cm. The collected data were firstly processed by the preliminary signal processors, during this stage, all the hyperbolic features indicating targets were transferred into focused images in the B-scans, and the B-scans were combined into 3D data set.

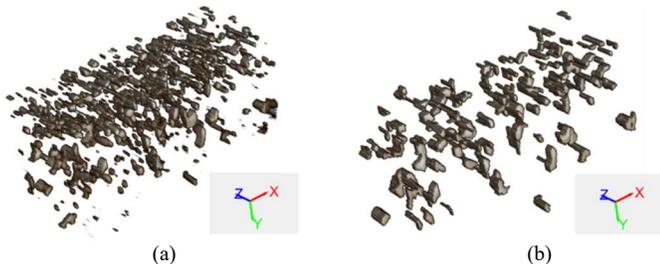

(a) (b)

Figure 2. 3D reconstruction of the tree roots. (a) without our algorithm, (b) with our algorithm. X is the scanning direction of each b-scan, Y is the depth underground, Z is the direction of tree roots extension.

Results that were processed without/with our algorithm are shown in Fig. 2(a)(b). After processing with our algorithm, reflections that are of no interest are ignored in Fig. 2(b), the tree roots reflections are clearer to recognize in Fig. 2(b) than that in Fig. 2(a).

Parts of the tree roots system that were dug out and their corresponding reconstructed image are shown in Fig.3 (a)(b). Fig. 3(b) is a cluster generated via the algorithm. Two images agree with each other, the extension trend of the roots is illustrated well by the result of our algorithm in Fig. 3(b).

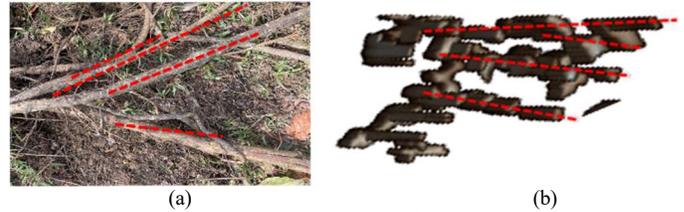

(a) (b)

Figure 3. Comparison between (a) ground truth, (b) extracted reconstruction results of tree roots system.

### IV. CONCLUSION AND FUTURE WORK

In this paper, an algorithm for the tree roots recognition by clustering technique has been introduced. The SSC algorithm is based on connecting focused image regions from adjacent slices of the 3D data set. It can cluster the separated focused signatures that are indicating targets in different B-scan slices. By doing so, the noises or reflections from targets that are not of interest can be clustered and removed from the original image. The real testing data result validates that our algorithm is effective in identifying the roots in noisy 3D GPR data. However, we believe that this SCC algorithm, which we are now investigating, can be further studied and work together with machine learning techniques, to classify different underneath targets, such as rocks, roots, pipes, etc.